\documentclass[preprint,review,12pt]{elsarticle}

\usepackage{amssymb}
\usepackage{longtable}

\begin{document}

\begin{frontmatter}



\title{Production of $\Sigma^{\pm}\pi^{\mp}pK^+$ in p+p reactions at 3.5 GeV beam energy}



\author[6]{G.~Agakishiev}
\author[3]{A.~Balanda}
\author[17]{D.~Belver}
\author[6]{A.~Belyaev}
\author[8]{J.C.~Berger-Chen}
\author[2]{A.~Blanco}
\author[9]{M.~B\"{o}hmer}
\author[15]{J.~L.~Boyard}
\author[17]{P.~Cabanelas}
\author[17]{E.~Castro}

% ...
\address[2]{LIP-Laborat\'{o}rio de Instrumenta\c{c}\~{a}o e F\'{\i}sica Experimental de Part\'{\i}culas , 3004-516~Coimbra, Portugal}
\address[3]{Smoluchowski Institute of Physics, Jagiellonian University of Cracow, 30-059~Krak\'{o}w, Poland}
% ...
\address[6]{Joint Institute of Nuclear Research, 141980~Dubna, Russia}
% ...
\address[8]{Excellence Cluster 'Origin and Structure of the Universe' , 85748~Garching, Germany}
\address[9]{Physik Department E12, Technische Universit\"{a}t M\"{u}nchen, 85748~Garching, Germany}
% ...
\address[15]{Institut de Physique Nucl\'{e}aire (UMR 8608), CNRS/IN2P3 - Universit\'{e} Paris Sud, F-91406~Orsay Cedex, France}
% ...
\address[17]{LabCAF. Dpto. F\'{\i}sica de Part\'{\i}culas, Univ. de Santiago de Compostela, 15706~Santiago de Compostela, Spain} 
% ...
\address[c]{Also at ExtreMe Matter Institute EMMI, 64291~Darmstadt, Germany}
% ...

\cortext[corr1]{laura.fabbietti@ph.tum.de}
\cortext[corr2]{johannes.siebenson@ph.tum.de}

\begin{abstract}
We study the production of $\Sigma^{\pm}\pi^{\mp}pK^+$ particle quartets in p+p reactions at 3.5 GeV kinetic beam energy. The data were taken with the HADES experiment at GSI. This report evaluates the contribution of resonances like $\Lambda(1405)$, $\Sigma(1385)^0$, $\Lambda(1520)$, $\Delta(1232)$, $N^*$ and $K^{*0}$ to the $\Sigma^{\pm} \pi^{\mp} p K^+$ final state. The resulting simulation model is compared to the experimental data in several angular distributions and it shows itself as suitable to evaluate the acceptance corrections properly.

\end{abstract}


\end{frontmatter}


\section{Introduction}
\label{}
For already half a century the $\Lambda(1405)$ is a well known resonance with strangeness $S=-1$, Isospin $I=0$ and spin $\frac{1}{2}$. Even though its four star character suggests a good understanding of this baryon, its inner structure is still a topic of investigation. Indeed it is difficult to describe the $\Lambda(1405)$ as a three quark baryon, as it is lighter than its nucleon partner, the $N^*(1535)$. Also the large mass difference to the $\Lambda(1520)$ can not be understood in terms of spin-orbital coupling \cite{Hyodo:2011ur}. With the mass of the $\Lambda(1405)$ lying slightly below the $\bar{K}N$ threshold another picture of this particle was established. From the analysis of the $\bar{K}N$ scattering length Dalitz and Tuan predicted the $\Lambda(1405)$ in 1959 \cite{Dalitz1959dn,Dalitz1959dq}, already two years before its experimental discovery.
Nowadays the $\Lambda(1405)$ is described in a coupled channel approach based on chiral dynamics \cite{Ikeda:2011dx}. Here this baryon is generated dynamically as an interference of two states, a $K^-p$ bound state and a $\Sigma\pi$ resonance. However, this two pole structure cannot be observed directly in the $\Sigma\pi$ invariant mass spectrum, as the $\Sigma\pi$ pole is located far in the imaginary part of the complex energy plane. \\
With these predictions, the structure of the $\Lambda(1405)$ is interesting in terms of a deeper understanding of the kaon-nucleon interaction. Experimental data are available for $\pi^-$+p \cite{Chao:1973sa}, $K^-$+p \cite{Hemingway:1984pz} and $\gamma$+p \cite{Moriya:2011af} reactions. First results on p+p data were reported in \cite{Zychor:2007gf}. But only the neutral decay channel ($\Lambda(1405)\rightarrow\Sigma^0\pi^0$) was  investigated. We also study p+p reactions and concentrate on the charged decay channels ($\Lambda(1405)\rightarrow\Sigma^{\pm}\pi^{\mp}$). In order to extract precisely the spectral function of the $\Lambda(1405)$, all the reactions that contribute to the $\Sigma^{\pm}\pi^{\mp}pK^+$ particle quartet have to be considered. This includes the production of resonances like $K^{0*}$, $N^*$ and $\Delta^{++}(1232)$. A simplified model, assuming only an incoherent sum of these contributions is finally used to describe the experimental data and extract the acceptance corrections. This model reproduces the experimental data for many kinematical variables.\\
The analyzed data were taken with the \textbf{H}igh  \textbf{A}cceptance  \textbf{D}i-\textbf{E}lectron  \textbf{S}pectrometer (HADES) \cite{Agakishiev:2009am} at GSI in Darmstadt, Germany. In this beam time a proton beam of 3.5 GeV kinetic energy was incident on a liquid hydrogen target and a total statistic of about 1.2 billion events was collected.   
                

\section{Data analysis}
\subsection{Evaluation of resonances contributions}
\begin{figure}[h]
	\centering
		\includegraphics[width=1.00\textwidth]{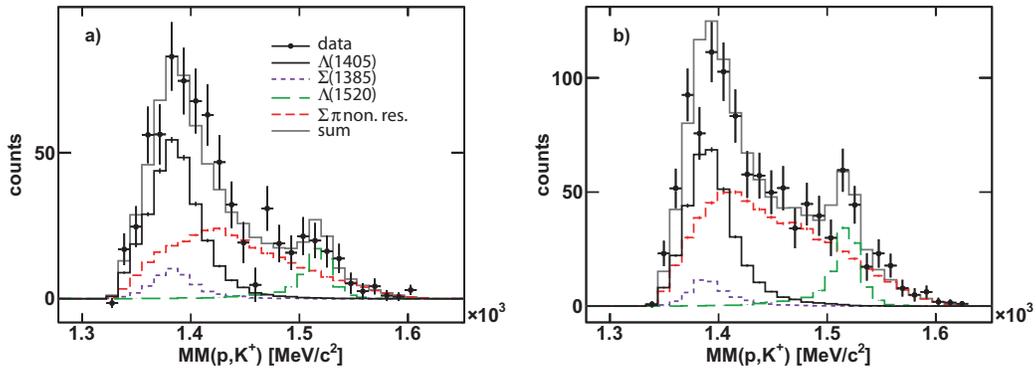}
	\caption{(Color online) Missing mass distribution of proton and $K^+$ for the two different subsamples within the HADES acceptance. Panel a) for events showing an intermediate $\Sigma^+$ and panel b) for events showing an intermediate $\Sigma^-$. Experimental data (black dots) are compared to simulations. See text for details.}
	\label{fig:LA1405_PlusNonRes}
\end{figure}

\begin{figure}[h]
	\centering
		\includegraphics[width=1.00\textwidth]{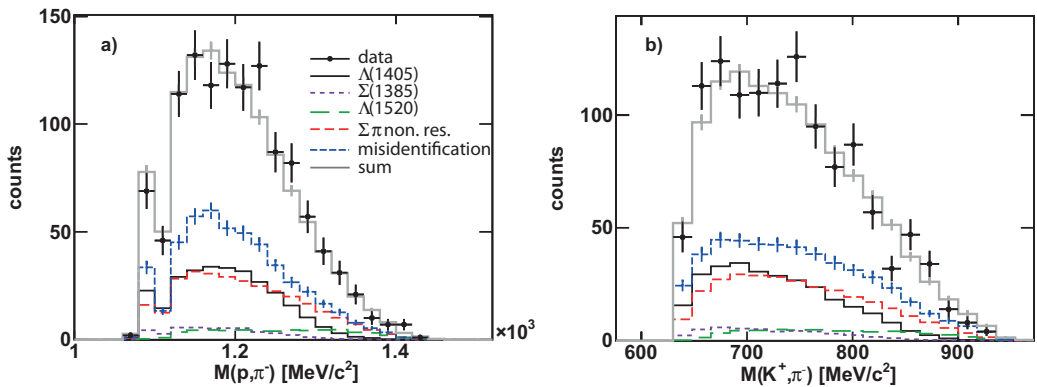}
	\caption{(Color online) a) Invariant mass of proton and $\pi^-$  and b) invariant mass of $K^+$ and $\pi^-$ for the subsample with an intermediate $\Sigma^+$ hyperon. The spectra show the results within the HADES acceptance.}
	\label{fig:InvMass_p_pim_K_pim}
\end{figure}

\begin{figure}[h]
	\centering
		\includegraphics[width=1.00\textwidth]{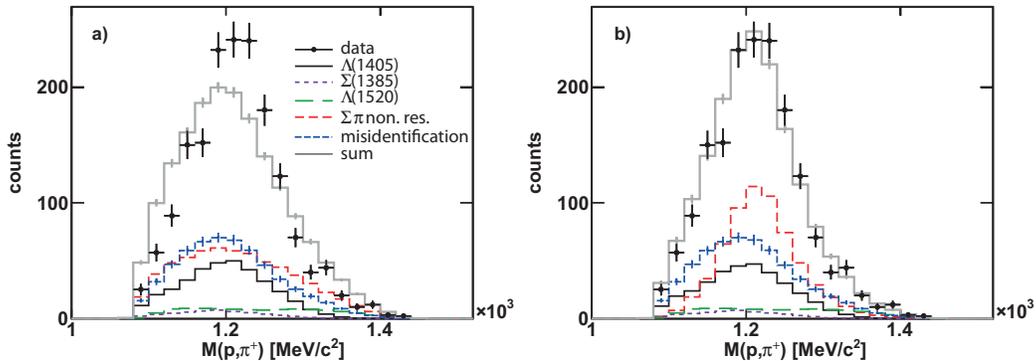}
	\caption{(Color online) Invariant mass of the proton and the $\pi^+$ within the HADES acceptance. Panel a): only channel 7 is used for the $\Sigma^-\pi^+$ non-resonant part. Panel b): according to a $\chi^2$ minimization, only channel 8 is used for the $\Sigma^-\pi^+$ non-resonant part.}
	\label{fig:InvMass_p_pip_both}
\end{figure}

The presented analysis concentrates on the production of the $\Lambda(1405)$ together with a proton and a $K^+$ followed by the decay of the $\Lambda(1405)$ into $\Sigma^{\pm}\pi^{\mp}$:
\begin{eqnarray}\label{rec:L1405}
p+p\rightarrow\Lambda(1405)+p+K^+\rightarrow(\Sigma^{\pm}\pi^{\mp})+p+K^+\rightarrow((n\pi^{\pm})\pi^{\mp})+p+K^+
\end{eqnarray}
The general analysis steps to extract the $\Lambda(1405)$ signal are presented in detail in \cite{Siebenson:2010hh,Siebenson:2010zz}. These steps consist in identifying the four charged final state particles ($p,K^+,\pi^+,\pi^-$) and the reconstruction of the neutron via the missing mass to the four particles. The neutron component can be enhanced by an appropriate cut on the corresponding missing mass. After this selection, the $\Sigma^+$ and $\Sigma^-$ hyperons are reconstructed via the missing mass of $p,K^+,\pi^-$ or $p,K^+,\pi^+$, respectively. By extracting the hyperon signals in the two spectra, the data sample is further purified, and at the same time it is divided into two subsamples. One subsample consists mainly of events with an intermediate $\Sigma^+$ hyperon, whereas the other subsample contains mainly events with an intermediate $\Sigma^-$ signal. For these two samples the missing mass spectrum of the proton and the $K^+$ ($MM(p,K^+)$), where the $\Lambda(1405)$ is expected to show up, is studied separately, see fig.~\ref{fig:LA1405_PlusNonRes}. 
To show the pure physical signal, in both pictures the misidentification background is already subtracted. The treatment of this misidentification background is discussed extensively in \cite{Agakishiev:2011qw}. The data (black dots) are compared to a sum of simulations. The strengths of the different contributions were determined by a simultaneous fit to four different observables, namely the two missing masses in fig.~\ref{fig:LA1405_PlusNonRes} and the two $p,K^+,\pi^{\mp}$ missing mass spectra where the missing mass of $\Sigma^+$ and $\Sigma^-$ are visible. Details about the fitting procedure to the four spectra can be found in \cite{Siebenson:2010hh}.
To give a full description of the experimental data, several contributions have to be taken into account in the simulation. A list of considered channels with particles in the same final state as in reaction (\ref{rec:L1405}) ($p,K^+,\pi^+,\pi^-$) is given in table~\ref{tab:Tabel1}. 
\begin{table}[h]
\begin{longtable}{||l|l|l||}\label{tab:Tabel1}
Channel & $p+p\rightarrow$ & Category\\ \hline
\endhead
1 & $\Lambda(1405)+p+K^+\rightarrow (\Sigma^{\pm}\pi^{\mp})+p+K^+$ & \\ \cline{1-2}
2 & $\Sigma(1385)^0+p+K^+\rightarrow(\Sigma^{\pm}\pi^{\mp})+p+K^+$ & $\Sigma\pi$ resonant\\ \cline{1-2}
3 & $\Lambda(1520)+p+K^+\rightarrow(\Sigma^{\pm}\pi^{\mp})+p+K^+$ & \\ \hline
4 & $\Sigma^++\pi^-+p+K^+$ & \\ \cline{1-2}
5 & $\Sigma^++K^++\Delta^0(1232)/N(1440)\rightarrow \Sigma^++K^++(p\pi^-)$ & $\Sigma^+\pi^-$ non-resonant\\ \cline{1-2}
6 & $\Sigma^++p+K^{*0}\rightarrow \Sigma^++p+(K^+\pi^-)$ & \\ \hline
7 & $\Sigma^-+\pi^++p+K^+$ & \\ \cline{1-2}
8 & $\Sigma^-+K^++\Delta^{++}(1232)\rightarrow \Sigma^-+K^++(p\pi^+)$ & $\Sigma^-\pi^+$ non-resonant\\ \hline
9 & $\Lambda+\pi^++n+K^+$ & \\ \cline{1-2}
10& $K^{0}_{S}+p+n+K^+$ & \\ \hline \hline 
\end{longtable}
\caption{Reactions taken into account for the analysis. The channels are classified into two main categories. See text for details.}
\end{table}
The channels 9 and 10 can be rejected from the data sample, as demonstrated in \cite{Siebenson:2010hh}. The other channels, however, contain all the same final and intermediate state particles and can therefore contribute to the data in fig.~\ref{fig:LA1405_PlusNonRes}. They are classified into two categories:
\begin{itemize}
\item ``$\Sigma\pi$ resonant'' are all channels, where the $\Sigma$ and the $\pi$ stem from the same mother particle. They should be visible as resonances in both $MM(p,K^+)$ spectra of fig.~\ref{fig:LA1405_PlusNonRes}. 
\item ``$\Sigma^+\pi^-$ ($\Sigma^-\pi^+$) non resonant'' are channels which have a $\Sigma^+\pi^-$ ($\Sigma^-\pi^+$) pair as an intermediate state, but the two particles are not stemming from a common mother particle. These channels give a broad, phase space like distribution in the spectra of fig.~\ref{fig:LA1405_PlusNonRes}. The ``$\Sigma^+\pi^-$ non resonant'' channels can only contribute to fig.~\ref{fig:LA1405_PlusNonRes} a), whereas the ``$\Sigma^-\pi^+$ non resonant'' channels can only give significant contribution to fig.~\ref{fig:LA1405_PlusNonRes} b).  
\end{itemize} 
Indeed, clear peak structures around 1400 MeV/c$^2$ and 1500 MeV/c$^2$ can be observed in both spectra of fig.~\ref{fig:LA1405_PlusNonRes}. They are attributed to the channels 1-3. The $\Lambda(1520)$ (green histograms) is well separated from the $\Lambda(1405)$ mass area and can therefore be isolated. However, the $\Sigma(1385)^0$ (violet histograms) overlaps completely with this area, and it is impossible to separate the $\Lambda(1405)$ and the $\Sigma(1385)^0$ in this data sample. Only in the neutral decay channels the two resonances show different properties ($\Lambda(1405)\rightarrow\Sigma^0\pi^0$, BR 33.3$\%$ and $\Sigma(1385)^0\rightarrow\Lambda\pi^0$, BR 88$\%$). This allows to analyze them independently. The obtained results for the p+p data at $E_{kin}=3.5$ GeV are reported in \cite{Epple:2011, EppleFabbietti:2011} and yield a cross section ratio of $\sigma_{\Lambda(1405)}/\sigma_{\Sigma(1385)^0}\approx 1$. This value is used as an external constraint for the analysis presented here. It gives the contributions of $\Lambda(1405)$ and $\Sigma(1385)^0$ shown in fig.~\ref{fig:LA1405_PlusNonRes}. 
For the simulation of the $\Lambda(1405)$ a Breit-Wigner distribution (black histograms) was used. However, for a good agreement between simulation and experiment, the Breit-Wigner had to be simulated with a pole mass of around 1385 MeV/c$^2$, which then results in the solid gray histograms.\\
To describe the spectra in fig.~\ref{fig:LA1405_PlusNonRes} completely, also phase space like distributions (red histograms), coming from the ``$\Sigma\pi$ non-resonant'' channels, are needed. A priori it is not clear to which extent the different channels in table~\ref{tab:Tabel1} contribute, as the spectra in fig.~\ref{fig:LA1405_PlusNonRes} are not sensitive to this information. Therefore, other observables have to be studied.\\
Fig.~\ref{fig:InvMass_p_pim_K_pim} concentrates on the subsample with an intermediate $\Sigma^+$ hyperon. It shows the same data set as in fig.~\ref{fig:LA1405_PlusNonRes} a), but before subtracting the misidentification background (blue histograms). Fig.~\ref{fig:InvMass_p_pim_K_pim} a) displays the invariant mass distribution of the proton and the $\pi^-$ ($M(p,\pi^-)$), where possible $\Delta$ and $N^*$ resonances should appear. For extracting a possible contribution of a $K^{*0}$, the invariant mass distribution of the $K^+$ and the $\pi^-$ ($M(K^+,\pi^-)$) is shown in fig.~\ref{fig:InvMass_p_pim_K_pim} b).  
Compared to the experimental data are simulations, where the scaling of the different channels is known from the simultaneous fit mentioned above. For the ``$\Sigma^+\pi^-$ non-resonant'' contribution only channel 4 is included. This assumption gives already a rather good description of the data. As  indications neither of $\Delta$/$N^*$ nor of $K^{*0}$ resonances are visible, only this channel 4 is used in the further analysis. Indeed, also the ``$\Sigma^+\pi^-$ non-resonant'' part shown in fig.~\ref{fig:LA1405_PlusNonRes} contains only this channel. However, possible contributions due to the channels 5 and 6 can not be excluded completely by this analysis. For example the production of a $K^{*0}(892)$ via channel 6 is only slightly above threshold and thus the cross section might be just too small to see a clear contribution to fig.~\ref{fig:LA1405_PlusNonRes}~b).   \\
To identify the different contributions to the ``$\Sigma^-\pi^+$ non-resonant'' part, the invariant mass of the proton and the $\pi^+$ ($M(p,\pi^+)$) is studied in fig.~\ref{fig:InvMass_p_pip_both}. 
Only the data subsample of fig.~\ref{fig:LA1405_PlusNonRes} b) with an intermediate $\Sigma^-$ hyperon is investigated. The misidentification background is not subtracted. Panel a) of fig.~\ref{fig:InvMass_p_pip_both} shows the result where the ``non resonant'' simulations contain only channel 7. The scaling factors for the different channels are again known from the fitting procedure. The data show an enhanced structure, which can not be described by the simulations. As a comparison, fig.~\ref{fig:InvMass_p_pip_both} b) shows exactly the same data, but now including channel 7 as well as channel 8 into the simulations. The relative contribution of these two channels is a free parameter, which is obtained by a $\chi^2$ fit to the experimental data points in fig.~\ref{fig:InvMass_p_pip_both}. The fit results in a negligible contribution of channel 7. With the inclusion of the $\Delta^{++}(1232)$ the experimental data can be described rather well. Due to this result, it is concluded that the ``$\Sigma^-\pi^+$ non-resonant'' contribution is dominated by  $\Delta^{++}(1232)$ production. Therefore, only the channel 8 is used in the simulation, which is already taken into account in fig.~\ref{fig:LA1405_PlusNonRes} b). \\
With the presented analysis a simulation model with several contributions is obtained, which gives reliable descriptions of the observables investigated so far. However, the goal of this analysis is to understand and to describe the experimental data within the full HADES acceptance. This asks for acceptance corrections. For this purpose it might be not sufficient to study only invariant mass distributions, as they are not very sensitive to e.g. angular distributions of the produced particles. Therefore, angular distributions in the Center-Mass System (CMS), Gottfried-Jackson system and helicity system are studied in the next part. Detailed information about the properties of these frames and the corresponding angular distributions can be found in \cite{Agakishiev:2011qw,AbdelBary:2010pc}.

\subsection{Angular distributions}
\begin{figure}[h!]
	\centering
		\includegraphics[width=0.99\textwidth]{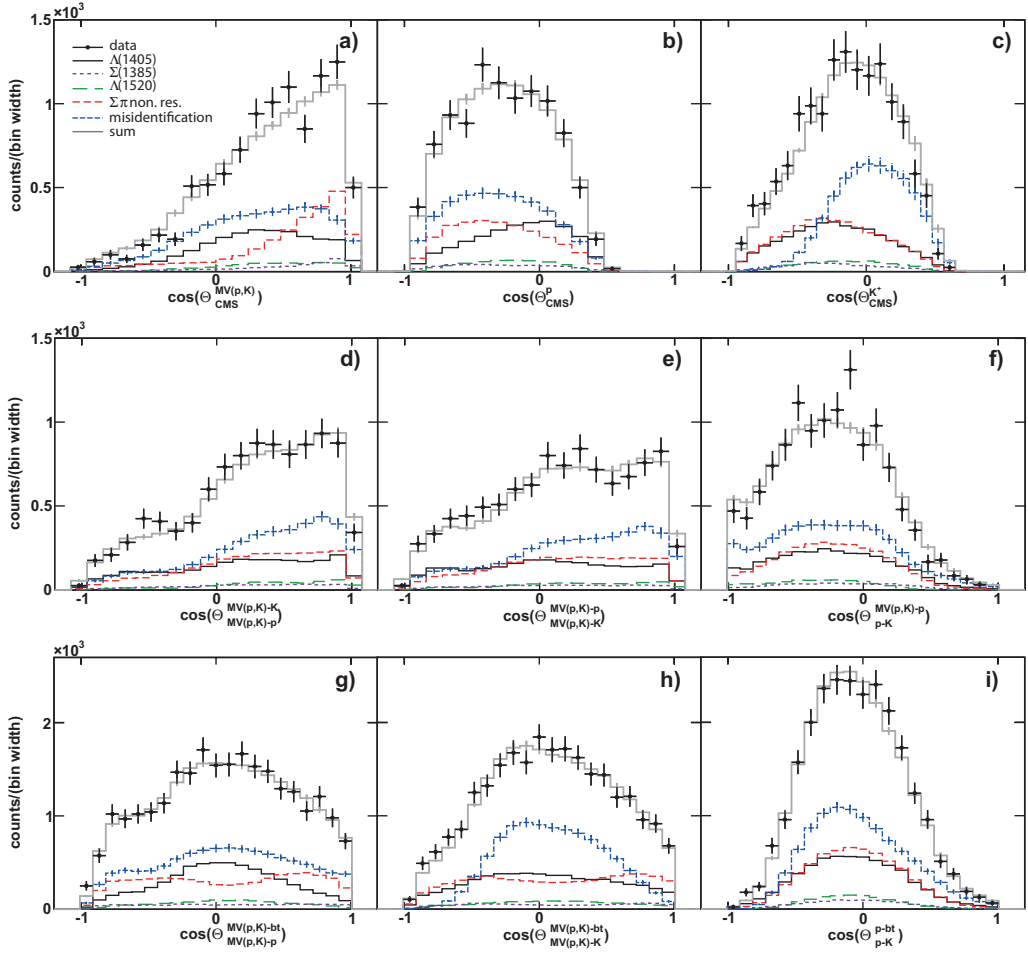}
	\caption{(Color online) Angular distributions within the HADES acceptance for events with an intermediate $\Sigma^+$ hyperon (top raw: distribution of $MV(p,K)$, $p$ and $K^+$ in the CMS, middle raw: helicity angles of $MV(p,K)$, $p$ and $K^+$, bottom raw: Gottfried-Jackson angles of $MV(p,K)$, $p$ and $K^+$.)}
	\label{fig:Angular_dist_SigmaP}
\end{figure}

\begin{figure}[h!]
	\centering
		\includegraphics[width=0.99\textwidth]{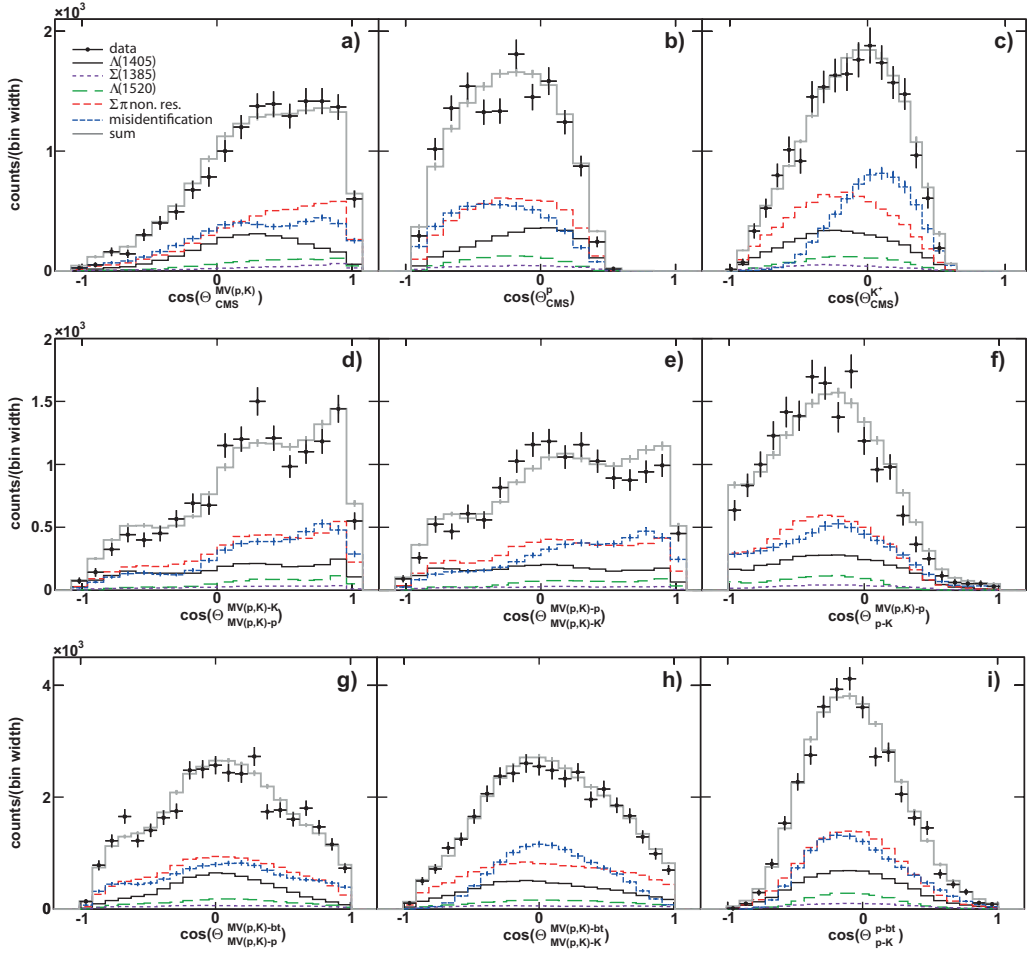}
	\caption{(Color online) Angular distributions within the HADES acceptance for events with an intermediate $\Sigma^-$ hyperon (top raw: distribution of $MV(p,K)$, $p$ and $K^+$ in the CMS, middle raw: helicity angles of $MV(p,K)$, $p$ and $K^+$, bottom raw: Gottfried-Jackson angles of $MV(p,K)$, $p$ and $K^+$.)}
	\label{fig:Angular_dist_SigmaM}
\end{figure}

Starting point is again reaction (\ref{rec:L1405}). Here, three particles are produced in the entrance channel ($\Lambda(1405)$, $p$ and $K^+$). The momentum of the possible $\Lambda(1405)$ is reconstructed via the missing four-vector to the proton and the $K^+$ ($MV(p,K)$). It is clear from the results above that this hypothetical particle does not always refer to a $\Lambda(1405)$, but can also stem from all other channels of table~\ref{tab:Tabel1}. Fig.~\ref{fig:Angular_dist_SigmaP} and \ref{fig:Angular_dist_SigmaM} show all angles between the three momenta in the three different frames for the subsample with an intermediate $\Sigma^+$ or $\Sigma^-$ hyperon, respectively. 
The nomenclature is the following:
\begin{itemize}
\item $\theta^{A}_{CMS}$: angle between particle A and the beam(target) direction in the Center-Mass System.
\item $\theta^{A-B}_{A-C}$: angle between particle A and particle B in the reference frame where particle A and particle C are going back to back and have equal momenta (helicity angle).
\item $\theta^{A-bt}_{A-C}$: angle between particle A and the beam/target-type protons (bt) in the reference frame where particle A and particle C are going back to back and have equal momenta (Gottfried-Jackson angle).
\end{itemize}
By permutation of particles A,B and C nine different observables are obtained, where some of them are not kinematically independent. 
The different panels in fig. \ref{fig:Angular_dist_SigmaP} and \ref{fig:Angular_dist_SigmaM} show the comparison between experimental data and simulations. A reasonable agreement could not be obtained by using only phase space simulations for the different channels of tab. \ref{tab:Tabel1}. The production of the $\Sigma(1385)^+$ in the CMS was found to be rather anisotropic \cite{Agakishiev:2011qw}. The production of the $\Sigma(1385)^0$ is assumed to show the same behavior. Therefore, the simulation of channel 2 was folded with an angular distribution in $cos\left(\theta^{MV(p,K)}_{CMS}\right)$. 
Additionally, the data sets in fig. \ref{fig:Angular_dist_SigmaP},a) and \ref{fig:Angular_dist_SigmaM},a) were subdivided into several angular regions of $\theta^{MV(p,K)}_{CMS}$. These subsamples were analyzed independently. In this way it was found that the $\Lambda(1405)$ and $\Lambda(1520)$ seem to be produced rather isotropically in $cos\left(\theta^{MV(p,K)}_{CMS}\right)$, whereas the ``$\Sigma^+\pi^-$ ($\Sigma^-\pi^+$) non resonant'' channels, namely channel 4 and 8, show an anisotropic behavior. This anisotropy is included by folding the simulation of these channels in $cos\left(\theta^{MV(p,K)}_{CMS}\right)$ with the corresponding distributions. The resulting distributions are included in fig. \ref{fig:Angular_dist_SigmaP} and \ref{fig:Angular_dist_SigmaM}. A reasonable agreement in all 18 pictures is obtained.   
    
\section{Summary}
We studied the production of the $\Sigma^{\pm}\pi^{\mp}pK^+$ particle quartets in p+p reactions. It was possible to describe the experimental missing mass distribution of the proton and $K^+$ by a sum of different simulations, including $\Lambda(1405)$, $\Sigma(1385)^0$, $\Lambda(1520)$ and ``non-resonant $\Sigma\pi$'' production. Studying several invariant mass spectra, the ``non-resonant $\Sigma^+\pi^-$'' production seems to stem mainly from the reaction $p+p\rightarrow\Sigma^++\pi^-+p+ K^+$ and no clear indication of intermediate resonances like $N^*/\Delta^0$ ($p+p\rightarrow N^*/\Delta^0+\Sigma^++K^+$) or $K^{0*}$ ($p+p\rightarrow\Sigma^++p+K^{*0}$) could be seen. However, the ``non-resonant $\Sigma^-\pi^+$'' production turned out to come exclusively from the reaction $p+p\rightarrow\Delta^{++}(1232)+\Sigma^-+K^+$. Furthermore, by including an anisotropic production of the $\Sigma(1385)^0$ channel as well as of the ``non-resonant $\Sigma\pi$'' channels, our simulations can describe the measured data for several angular distributions. The overall good agreement between experimental data and simulations is a necessary precondition to use the obtained simulation model for acceptance and efficiency corrections and to finally extract cross sections for the different channels.     

\section{Acknowledgements}
The author gratefully acknowledge support
from the TUM Graduate School.\\
The following funding are acknowledged.
LIP Coimbra,
Coimbra (Portugal): PTDC/FIS/113339/2009,
SIP JUC Cracow, Cracow (Poland): NN202286038, NN202198639, HZ Dresden-Rossendorf, Dresden
(Germany): BMBF 06DR9059D, TU Muenchen,
Garching (Germany) MLL Muenchen DFG EClust:
153 VH-NG-330, BMBF 06MT9156 TP5 TP6, GSI
TMKrue 1012, GSI TMFABI 1012, NPI AS CR,
Rez (Czech Republic): MSMT LC07050, GAASCR
IAA100480803, USC - S. de Compostela, Santiago de
Compostela (Spain): CPAN:CSD2007-00042, Goethe
Univ. Frankfurt (Germany): HA216/EMMI, HIC
for FAIR (LOEWE), BMBF06FY9100I, GSI F\&E01,
CNRS/IN2P3 (France).






\end{document}